\documentclass[12pt]{article}

\usepackage{color}
\usepackage{amsmath}
\usepackage{graphicx}
\usepackage{amssymb}
\usepackage{bm}

\numberwithin{equation}{section}

\title{Renormalization group formalism for incompressible Euler equations and the blowup problem}

\author{Alexei A. Mailybaev\footnote{Instituto 
Nacional de Matem\'atica Pura e Aplicada -- IMPA, Estrada Dona Castorina 110, 22460-320 Rio de Janeiro, RJ, Brazil. 
Phone:~+55\,21\,2529\,5070, Fax:~+55\,21\,2529\,5075, 
E-mail: alexei@impa.br.} \footnote{Institute 
of Mechanics, Lomonosov Moscow State University, Russia}}

\begin{document}

\maketitle

\begin{abstract}
The paper discusses extensions of 
the renormalization group (RG) formalism for 
3D incompressible Euler equations, which can be used for 
describing singularities developing in finite (blowup) 
or infinite time from smooth initial conditions of finite energy. 
In this theory, time evolution is substituted by the equivalent 
evolution for renormalized solutions governed by the RG equations.
A fixed point attractor of the RG equations, if it exists, 
describes universal self-similar form of observable singularities. 
This universality provides a constructive criterion for interpreting 
results of numerical experiments.   
In this paper, renormalization schemes with multiple spatial scales 
are developed for the cases of power law and exponential scaling.
The results are compared with the numerical 
simulations of a singularity in incompressible Euler equations 
obtained by Hou and Li (2006) and Grafke et al. (2008). 
The comparison supports the conjecture of a singularity developing exponentially in infinite time and described by a multiple-scale 
self-similar asymptotic solution predicted by the RG theory. 
\end{abstract}



\maketitle

\section{Introduction}

The question of whether the incompressible Euler equations in 
three dimensions can develop a finite time singularity (blowup) from smooth initial conditions 
of finite energy is the long-standing open problem in fluid dynamics
\cite{chae2008incompressible,constantin2007euler,gibbon2008three,gibbon2008three2}. 
This question is of fundamental importance, as the 
blowup may be related to the onset of turbulence~\cite{holm2002transient} 
and to the energy transfer to small scales~\cite{eyink2006onsager}. 
Direct numerical simulations provide a powerful tool to probe the blowup. 
However, despite of large effort, we are still far from having a definite answer.  

There is a number of blowup and no-blowup criteria, which are useful in numerical 
simulations to detect a finite-time singularity. The widely used criterion 
is due to the Beale--Kato--Majda theorem~\cite{beale1984remarks}, which states 
that the time integral of maximum vorticity must explode at a singular point. 
Several criteria, which also use the direction of vorticity, are developed by 
Constantin \textit{et al.} \cite{constantin1996geometric}, 
Deng  \textit{et al.} \cite{deng2006improved,deng2005geometric} 
and Chae \cite{chae2007finite}. 
See also \cite{chae2008incompressible,hou2009blow,kuznetsov2003towards} and references therein. 

The history of numerical studies is summarized in \cite{gibbon2008three,hou2009blow,kerr2006computational}. 
Conclusions based on numerical 
simulations vary depending on initial conditions and numerical method. However, 
none of the results seem to be sufficiently convincing so far. 
Apparently, the success of numerical simulations requires 
further development of the theory.  

In this paper, we consider the renormalization group (RG) approach to the blowup problem. 
The RG method is famous to capture 
sophisticated critical phenomena characterized by scaling universality, e.g., 
critical phenomena in second-order phase transitions~\cite{wilson1974renormalization} 
and period-doubling route to chaos~\cite{feigenbaum1978quantitative}, 
see also the review in \cite{kadanoff2011}. 
There are various applications of this method to problems of fluid dynamics~\cite{eggers2009role}. 
Universality of self-similar blowup was explained in~\cite{dombre1998intermittency} 
for inviscid shell models of turbulence using the method, 
which is similar in spirit to the RG approach. Analogous universal self-similar 
blowup was observed in cascade models of the Euler equations~\cite{uhlig1997singularities}. 
The RG approach to the blowup problem for incompressible 
Euler equations was discussed in~\cite{greene1997evidence,greene2000stability,mailybaev2012} 
in the case of a single spatial scale. 
In this paper, we study extension of the RG method for multiple spatial scales 
in the cases of power law and exponential scaling. Note that the universality predicted by the RG theory provides a constructive criterion for interpretation of numerical results. 

We start by illustrating an idea of the RG method on the inviscid Burgers equation, 
where the blowup phenomenon is simple and well known. Here the RG equation is derived for solutions renormalized near a singularity. Fixed points of the 
RG equation correspond to self-similar blowup solutions. 
Existence of an attracting fixed point explains universal scaling of the 
blowup~\cite{eggers2009role,mailybaev2012}.

The RG equations for incompressible Euler equations are derived 
first for the case of a single spatial scale. Fixed points of the RG equations 
correspond to exact self-similar solutions. 
A part of this theory corresponding to finite time singularities with power law scaling
was considered in \cite{greene1997evidence,greene2000stability,mailybaev2012}. 
We extend the RG formalism to the case of singularities developing exponentially in infinite time. 
Note that exponential scaling in this problem was suggested 
in \cite{brachet1983small,brachet1992numerical} based on numerical results.

The main contribution of this paper is the development of the RG formalism with 
multiple spatial scales. Self-similar solutions with such scaling cannot be
exact solutions of the Euler equations. 
However, they may serve as asymptotic solutions. This is proved by introducing a 
special renormalization of the pressure term in the RG equations. 
Attracting fixed point solutions of the RG equations, if they exist, 
describe universal asymptotic form of observable flow singularities. 
Two types of scaling are considered, which correspond to finite time (blowup) 
and infinite time (exponential) singularities.    

The asymptotic forms of singularities provided by the RG theory are tested using the 
results of numerical simulations obtained by Hou and Li~\cite{hou2007computing,hou2008blowup} 
and Grafke \textit{et al.}~\cite{grafke2008numerical}. 
The comparison supports the conjecture of \cite{brachet1983small,brachet1992numerical} on exponential scaling of a singularity, which is consistent with a multiple-scale 
self-similar form of solution allowed by the RG theory. 

The paper is organized as follows. Section~\ref{secBurg} describes the RG theory 
for the inviscid Burgers equation. Section~\ref{secExp} describes the single-scale 
version of the RG formalism for the incompressible Euler equations. Section~\ref{secMS} 
develops the RG theory in the case of multiple scales. 
Section~\ref{secExp2} extends the results to the case of multiple-scale 
exponential singularities.
Section~\ref{secSin} compares the theory with known numerical results. 
Conclusion summarizes the contribution.  

\section{Attractor of renormalized Burgers equation}
\label{secBurg}

In this section we demonstrate the idea of the RG approach on a simple example of the inviscid Burgers equation
\begin{equation}
u_t +uu_x = 0,
\label{eq1}
\end{equation}
which has the well-known classical solution leading to a singularity in finite time (blowup). 
This example contains many features of the RG 
formalism for the incompressible Euler equations developed in the next sections.

Solution of equation (\ref{eq1}) is 
given implicitly by the method of characteristics as
\begin{equation}
u(x,t) = u_0(z),\quad 
x = z+u_0(z)t,
\label{Z.2}
\end{equation}
where $u_0(x)$ is a smooth initial condition at $t = 0$ and $z$ is an auxiliary variable. 
The blowup, $u_x \to \infty$, occurs when $\partial x/\partial z = 0$. 
Expressing this derivative from the second expression in (\ref{Z.2}) yields 
$1+\partial_z u_0(z)\, t = 0$. 
Therefore, one finds the time $t_b$ and position $x_b = z$ of the blowup from the condition 
\begin{equation}
t_b = \min_{z}\left(-\frac{1}{\partial_z u_0(z)}\right).
\label{Z.3}
\end{equation}
We assume that $u_b = u(x_b,t_b) = 0$ at the blowup point. This condition can always be satisfied by means of the Galilean transformation, 
which is a symmetry of (\ref{eq1}).

Following \cite{eggers2009role}, we introduce the variables 
\begin{equation}
x' = x-x_b,\quad t' = t_b-t, 
\label{Z.var}
\end{equation}
where $t' > 0$ is the time interval to the blowup, and 
consider the renormalized solutions defined as
\begin{equation}
u(x,t) = t'^{(a-1)}U(\xi,\tau),\quad \xi = x'/t'^{a},\quad \tau = -\log t'.
\label{Z.4}
\end{equation}
In the new variables, the blowup corresponds to $\tau \to \infty$. 
Equation for $U(\xi,\tau)$ is found from (\ref{eq1}), (\ref{Z.4}) as
\begin{equation}
U_\tau = (a-1)U-a\xi U_\xi-UU_\xi.
\label{Z.5}
\end{equation}
For $a = 3/2$, this equation has a stable fixed point solution $U(\xi,\tau) = U_*(\xi)$ given implicitly 
by \cite{eggers2009role}
\begin{equation}
\xi = -U_*-CU_*^3,\quad C > 0.
\label{Z.6}
\end{equation}
Moreover, one can show \cite{pomeau2008wave} that 
\begin{equation}
\lim_{\tau \to \infty} U(\xi,\tau) = U_*(\xi)
\label{Z.lim}
\end{equation}
for any blowup solution with nondegenerate minimum (\ref{Z.3}) and some constant $C$. Note 
that this result extends to general scalar conservation laws~\cite{mailybaev2012}.

\begin{figure}
\centering \includegraphics[width=0.44\textwidth]{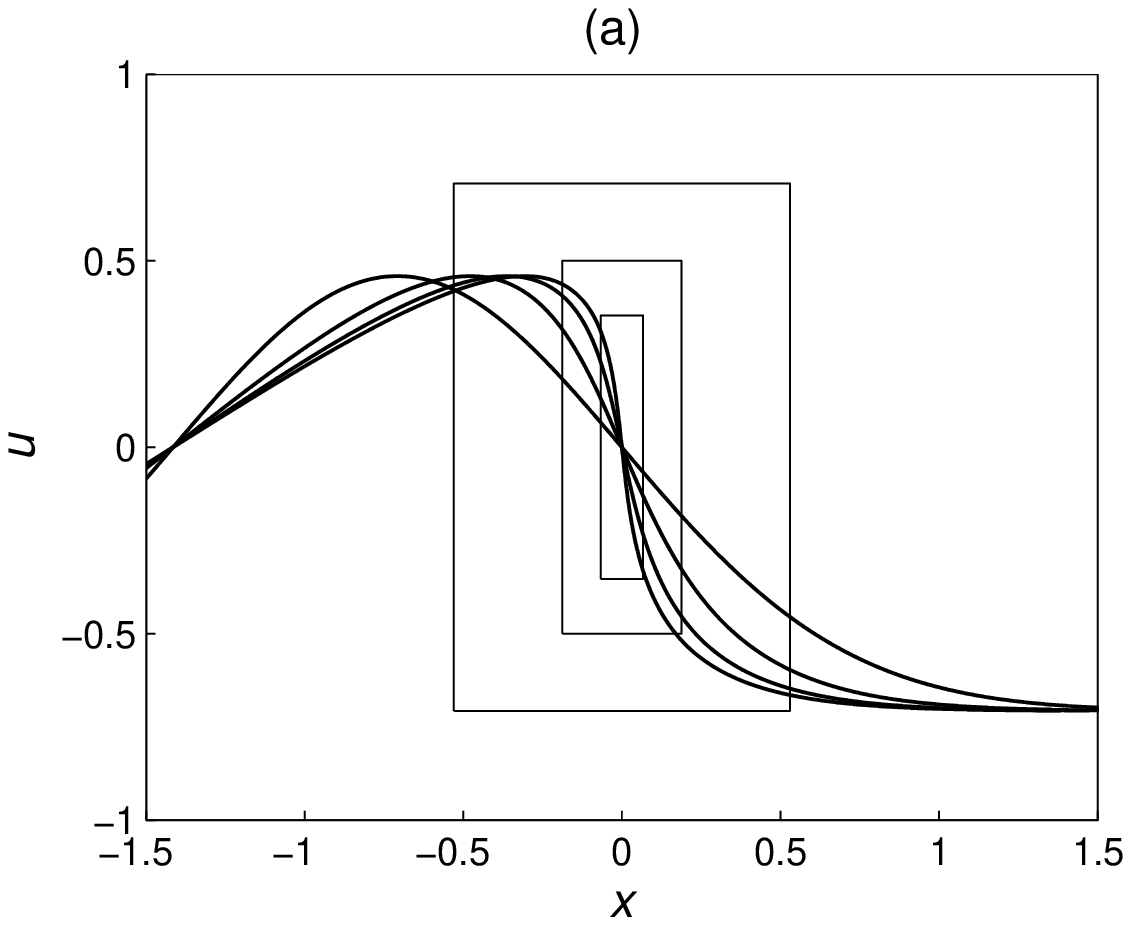}
\hspace{2mm}
\centering \includegraphics[width=0.44\textwidth]{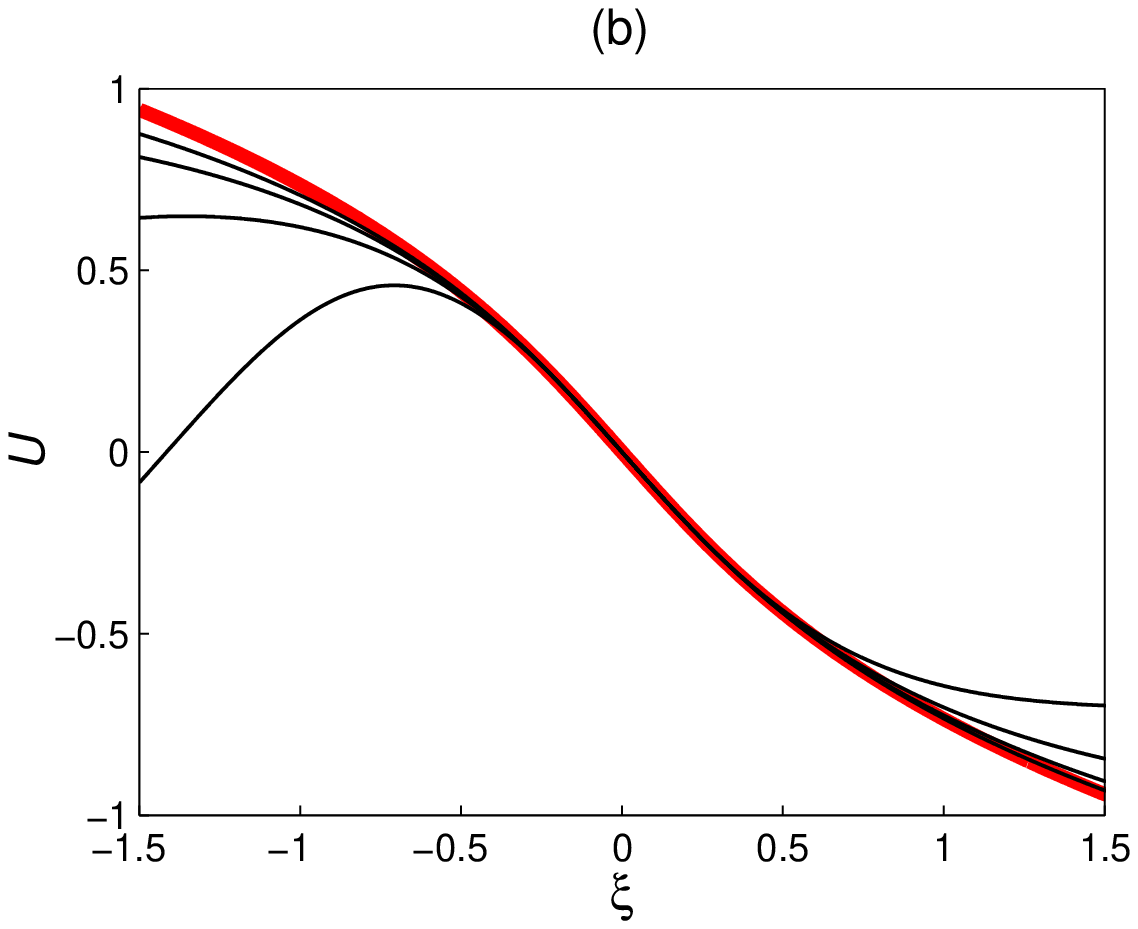}
\caption{(a) Solution of the inviscid Burgers equation at  
$t' = 1$, $1/2$, $1/4$, and $1/8$, where $t' = t_b-t$ is the time to blowup. 
Renormalization boxes, which scale as $x \sim t'^{3/2}$ and $u \sim t'^{1/2}$, 
are shown. (b) Renormalized solutions converge 
to the universal function $U_*(\xi)$, which is shown by the bold red curve.  } 
\label{fig3}
\end{figure}

Expressions (\ref{Z.4}) with $a = 3/2$ and (\ref{Z.lim}) imply
the asymptotic relation
\begin{equation}
u(x,t) \to t'^{1/2}U_*(x'/t'^{3/2}),\quad x' \sim t'^{3/2}, \quad t' \to 0.
\label{Z.10}
\end{equation}
An important observation is that this asymptotic relation is valid in the vanishingly small 
neighborhood $x' \sim t'^{3/2}$ of a singularity. For large $\xi = x'/t'^{3/2}$, 
using the approximation $U_* \approx -(\xi/C)^{1/3}$ 
following from (\ref{Z.6}) in (\ref{Z.10}) leads to the classical cubic-root expression $u(x,t) \approx -(x/C)^{1/3}$. 
Therefore, the wave profile $u(x,t)$ contains a universal self-similar ``core'' 
of the form (\ref{Z.10}) shrinking to a point as the blowup is approached. This ``core'' 
has the cubic root $x$-dependence for $x' \gg t'^{3/2}$, where it is ``glued'' 
to the rest of the solution. An example 
of convergence to the fixed point solution $U_*$ is shown in Fig.~\ref{fig3}.   

The scaling symmetry
\begin{equation}
u(x,t) \mapsto b_0u(x/b_0,t)
\label{Z.8}
\end{equation}
can be used to set the value $C = 1$ in (\ref{Z.6}). 
The choice of $t_b$, $x_b$ at the blowup and the condition $u_b = 0$ are 
important for the convergence to $U_*$ in (\ref{Z.lim}), 
as small errors in satisfying these conditions lead to instability \cite{eggers2009role}. 
The values of $x_b$, $t_b$ and $u_b$ can always be adjusted by using 
the symmetry transformations
\begin{equation}
u(x,t) \mapsto u(x+b_1,t), \quad
u(x,t) \mapsto u(x,t+b_2), \quad
u(x,t) \mapsto u(x+b_3t,t)-b_3.
\label{Z.7}
\end{equation}

\begin{figure}
\centering \includegraphics[width=0.45\textwidth]{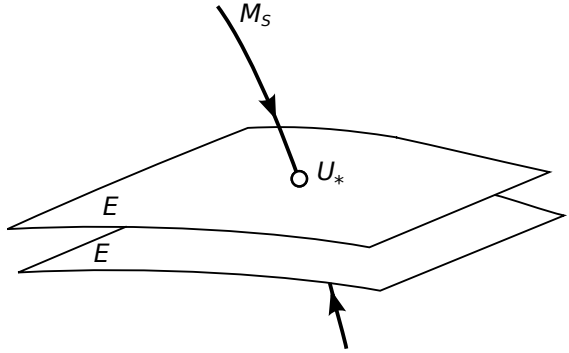}
\caption{Schematic local structure of the fixed point $U_*$ 
of the RG equation (\ref{Z.5}). $M_S$ represents the stable  manifold. 
Points of each surface $E$ correspond to solutions related by symmetries (\ref{Z.8}) and (\ref{Z.7}). 
Universal structure of the blowup is explained by the RG dynamics on the stable manifold.} 
\label{fig2}
\end{figure}

In summary, the blowup in the inviscid Burgers equation is related to the 
evolution of the renormalized function $U(\xi,\tau)$ 
near the fixed point $U_*(\xi)$ 
in functional space demonstrated schematically in Fig.~\ref{fig2}. 
The fixed point $U_*$ has a stable manifold $M_S$ of codimension 4. 
The four extra dimensions correspond 
to solutions related by symmetries (\ref{Z.8}) and (\ref{Z.7}), which determine 4-dimensional 
surfaces $E$ of equivalent solutions. Surfaces $E$ intersect $M_S$, 
so that one can move any point to the stable manifold by the symmetry transformation. 
This, in turn, implies universality of the limiting renormalized solution.

One can check that the shift $\tau \mapsto \tau+\Delta\tau$ in (\ref{Z.4}) 
induces renormalization in space-time. 
This shift can be viewed as an action 
of the renormalization operator 
\begin{equation}
U(\xi,\tau) \xrightarrow{\mathcal{R}_{\Delta\tau}} U(\xi,\tau+\Delta\tau).
\label{Z.9}
\end{equation}
Since the right-hand side of
(\ref{Z.5}) does not depend on $\tau$ explicitly, this operator generates 
a one-parameter renormalization group with $\mathcal{R}_{\tau_1+\tau_2} = \mathcal{R}_{\tau_1}
\mathcal{R}_{\tau_2}$. Existence of such a group is an important property underlying 
the above construction, similarly to other applications 
of the RG theory~\cite{wilson1974renormalization,feigenbaum1978quantitative,kadanoff2011}. 

\section{Exponential renormalization of incompressible Euler equations}
\label{secExp}

The Euler equations governing the flow of ideal incompressible fluid of unit density in three dimensional space $\mathbf{x} = (x_1,x_2,x_3)$ are 
\begin{equation}
\mathbf{u}_t+\mathbf{u}\cdot\nabla\mathbf{u} 
= -\nabla p,\quad \nabla\cdot \mathbf{u} = 0.
\label{X.1}
\end{equation}
We consider solutions $\mathbf{u}(\mathbf{x},t)$ with smooth initial conditions of finite energy, $\int d^3x \|\mathbf{u}\|^2 < \infty$. 

First, let us assume the finite time blowup, when the flow forms a singularity at $(\mathbf{x}_b,t_b)$. Using the Galilean transformation, one can set $\mathbf{u}_b = \mathbf{u}(\mathbf{x}_b,t_b) = 0$. Leray \cite{leray1934mouvement} suggested to consider self-similar solutions 
of the form 
\begin{equation}
\mathbf{u} = t'^{(a-1)}\mathbf{U}_*(\mathbf{x}'/t'^{a}),\quad
p = t'^{(2a-2)}P_*(\mathbf{x}'/t'^{a}),\quad a > 0,
\label{X.0}
\end{equation}
in the study of blowup, where 
\begin{equation}
\mathbf{x}' = \mathbf{x}-\mathbf{x}_b,\quad t' = t_b-t. 
\label{X.blow}
\end{equation}
Finite energy solutions of this form cannot be realized 
globally~\cite{chae2007nonexistence,chae2008incompressible}. However, the global existence is not required in the RG theory, 
since the self-similar expression is valid asymptotically in a vanishingly 
small neighborhood of a singularity, see (\ref{Z.10}).   

The renormalized solution is defined similarly to (\ref{Z.4}) as
\begin{equation}
\mathbf{u}(\mathbf{x},t) 
= t'^{(a-1)}\mathbf{U}({\bm \xi},\tau ),\quad 
p(\mathbf{x},t) 
= t'^{(2a-2)}P({\bm \xi},\tau ),\quad 
{\bm \xi} = \mathbf{x}'/t'^{a},\quad \tau = -\log t'.
\label{X.2}
\end{equation}
Substituting these expressions into (\ref{X.1}), one obtains the RG equations (see~\cite{greene1997evidence,greene2000stability})
\begin{equation}
\mathbf{U}_\tau
= (a-1)\mathbf{U}-(a{\bm\xi}+\mathbf{U})\cdot \nabla_{\xi}
\mathbf{U}-\nabla_{\xi}P, \quad 
\nabla_{\xi}\cdot\mathbf{U} = 0.
\label{X.3}
\end{equation}
A fixed point solution $\mathbf{U}_*({\bm\xi})$, $P_*({\bm\xi})$ satisfies the equations 
\begin{equation}
(a-1)\mathbf{U}_*-(a{\bm\xi}+\mathbf{U}_*)\cdot \nabla_{\xi}
\mathbf{U}_*-\nabla_{\xi}P_* = 0, \quad 
\nabla_{\xi}\cdot\mathbf{U}_* = 0,
\label{X.4}
\end{equation}
and determines the self-similar solution (\ref{X.0}). The vorticity computed for 
the velocity (\ref{X.0}) grows as ${\bm\omega} \sim 1/t'$, which agrees with the Beale--Kato--Majda theorem~\cite{beale1984remarks}.

Fixed point solutions 
describing the blowup are not known, but something can be said about their stability assuming that they exist \cite{greene2000stability}. 
Stable fixed points would describe observable blowup phenomena. 
Different types of attractors of the RG equations like, e.g., 
periodic solutions are also relevant for the blowup problem~\cite{pomeau2005unfinished}. 

The RG formalism can be extended to the case of a singularity developing exponentially in infinite time. Assuming that the solution is regular for all $t \ge 0$ (no finite time blowup), we consider the 
renormalized functions $\widetilde{\mathbf{U}}({\bm\xi},t)$ and $\widetilde{P}({\bm\xi},t)$ defined as
\begin{equation}
\mathbf{u}(\mathbf{x},t) 
= e^{-bt}\widetilde{\mathbf{U}}({\bm \xi},t),\quad 
p(\mathbf{x},t) 
= e^{-2bt}\widetilde{P}({\bm \xi},t),\quad 
{\bm \xi} = \mathbf{x}'/e^{-bt},\quad b > 0.
\label{X.5}
\end{equation}
Substituting these expressions into (\ref{X.1}), one obtains the RG equations 
\begin{equation}
\widetilde{\mathbf{U}}_t
= b\widetilde{\mathbf{U}}-(b{\bm\xi}
+\widetilde{\mathbf{U}})\cdot \nabla_{\xi}
\widetilde{\mathbf{U}}-\nabla_{\xi}\widetilde{P}, \quad 
\nabla_{\xi}\cdot\widetilde{\mathbf{U}} = 0.
\label{X.6}
\end{equation}
A fixed point solution $\widetilde{\mathbf{U}}_*({\bm\xi})$, 
$\widetilde{P}_*({\bm\xi})$ satisfies the equations 
\begin{equation}
b\widetilde{\mathbf{U}}_*-(b{\bm\xi}
+\widetilde{\mathbf{U}}_*)\cdot \nabla_{\xi}
\widetilde{\mathbf{U}}_*-\nabla_{\xi}\widetilde{P}_* = 0, \quad 
\nabla_{\xi}\cdot\widetilde{\mathbf{U}}_* = 0,
\label{X.7}
\end{equation}
and determines the self-similar solution  
\begin{equation}
\mathbf{u}(\mathbf{x},t) 
= e^{-bt}\widetilde{\mathbf{U}}(\mathbf{x}'/e^{-bt}),\quad 
p(\mathbf{x},t) 
= e^{-2bt}\widetilde{P}(\mathbf{x}'/e^{-bt})
\label{X.8}
\end{equation}
describing a singularity developing exponentially as $t \to \infty$. 
Note that one can always set $b = 1$ by time scaling.

The vorticity computed for 
the velocity (\ref{X.8}) remains finite, 
but second spatial derivatives of the velocity grow exponentially with time. 
Since vorticity grows in all numerical simulations of the incompressible Euler equations, it is unlikely that stable fixed points of (\ref{X.6}) exist. 
However, analogous exponential singularities with multiple scales described 
below seem 
to be good candidates for describing observable phenomena. 

We remark that equations (\ref{X.5})--(\ref{X.8}) follow from (\ref{X.0})--(\ref{X.4}) 
in the limit $a \to \infty$. For example, the fixed point equations (\ref{X.7}) 
are obtained from (\ref{X.4}) in the limit $a \to \infty$ after multiplication 
by $b^2/a^2$ and the substitution $\mathbf{U}_* = (a/b)\widetilde{\mathbf{U}}_*$, 
$P_* = (a/b)^2\widetilde{P}_*$. Similar relation is established for other expressions by 
taking $t' = 1-b\widetilde{t}/a$ so that 
\begin{equation}
t'^{a} = (1-b\widetilde{t}/a)^a \to e^{-b\widetilde{t}},\quad
\tau = -\log t' \to b\widetilde{t}/a
\label{X.9}
\end{equation}
as $a \to \infty$.

Recall that, in the incompressible Euler equations, the pressure is determined 
by the velocity field at the same time. The same, of course, 
is valid for the RG equations. Indeed,
computing divergence of both sides of the first expression in (\ref{X.6}), 
the left-hand side vanishes due to the incompressibility condition and the 
right-hand side yields Poisson's equation for $\widetilde{P}$.
Its solution is well determined if the vector field 
$\widetilde{\mathbf{U}}({\bm\xi},t)$ decays sufficiently fast as $\|{\bm\xi}\| \to \infty$. 
However, following the results of Section~\ref{secBurg}, we can expect that the fixed point solution 
$\widetilde{\mathbf{U}}_*({\bm\xi})$ of the RG equations is unbounded for large $\|\bm\xi\|$. 
In this case the problem (\ref{X.7}) for a fixed point is not well-posed. 
This inconsistency is removed in the multiple-scale RG theory 
considered in the next sections. 

\section{Multiple-scale RG formalism}
\label{secMS}

Self-similar expressions of the form (\ref{X.0}) or (\ref{X.8}) describe singularities with a single spatial scale, $x \sim t'^a$ or $x \sim e^{-bt}$. On the other hand,
numerical simulations of incompressible Euler equations available in the 
literature demonstrate very thin singular structures 
implying existence of at least two different scaling laws. 
For example, the two scales proposed in \cite{kerr2005velocity} are 
$x_1 \sim 1/\sqrt{t'}$ and $x_2 \sim 1/t'$. 
In this section, we generalize the RG formalism for the case of multiple-scale self-similar solutions.

Let us assume that solution $\mathbf{u}(\mathbf{x},t)$ with smooth initial condition of finite energy blows up in finite time at $(\mathbf{x}_b,t_b)$. Using the Galilean transformation, we set $\mathbf{u}_b = \mathbf{u}(\mathbf{x}_b,t_b) = 0$. 
Let us introduce the diagonal matrices
\begin{equation}
t'^{\mathbf{A}} 
= \mathrm{diag}\,(t'^{a_1},t'^{a_2},t'^{a_3}),\quad
\mathbf{A} = \mathrm{diag}\,(a_1,a_2,a_3),
\label{Y.1}
\end{equation}
which generalize the scaling $\mathbf{x}'\sim t'^a$ of Section~\ref{secExp} to the multiple-scale case.
In what follows, we assume 
\begin{equation}
0 < a_1 \le a_2 < a_3
\label{Y.0}
\end{equation}
with a single dominant power $a_3$.
The case of equal $a_2$ and $a_3$ can be analyzed similarly. 

The renormalized solution is defined as
\begin{equation}
\mathbf{u}(\mathbf{x},t) 
= t'^{(\mathbf{A}-\mathbf{I})}\mathbf{U}({\bm \xi},\tau ),\quad 
p(\mathbf{x},t) 
= t'^{(2a_3-2)}P({\bm \xi},\tau ),\quad 
{\bm \xi} = t'^{-\mathbf{A}}\mathbf{x}',\quad \tau = -\log t',
\label{Y.2}
\end{equation}
where $\mathbf{I}$ is the identity matrix.
The first relation written for each vector component has the form
\begin{equation}
u_j = t'^{(a_j-1)}U_j({\bm \xi},\tau ),\quad j = 1,2,3,\quad
{\bm\xi} = (x_1/t'^{a_1},x_2/t'^{a_2},x_3/t'^{a_3}),
\label{Y.2b}
\end{equation}
where three different scales are used for different space directions. 

Substituting (\ref{Y.2}) into (\ref{X.1}) yields
\begin{equation}
\mathbf{U}_\tau
= (\mathbf{A}-\mathbf{I})\mathbf{U}-(\mathbf{A}{\bm\xi}+\mathbf{U})\cdot \nabla_{\xi}
\mathbf{U}-\mathbf{C}\nabla_{\xi}P, \quad 
\nabla_{\xi}\cdot\mathbf{U} = 0,
\label{Y.3}
\end{equation}
where $\mathbf{C}$ is the diagonal matrix 
\begin{equation}
\mathbf{C}(\tau) = t'^{2a_3}t'^{-2\mathbf{A}} 
= e^{-2a_3\tau}e^{2\mathbf{A}\tau} = \mathrm{diag}\,(e^{2\tau(a_1-a_3)},e^{2\tau(a_2-a_3)},1).
\label{Y.4}
\end{equation}
Taking time derivative of (\ref{Y.4}) yields the equation
\begin{equation}
\mathbf{C}_\tau = 2(\mathbf{A}-a_3\mathbf{I})\mathbf{C}.
\label{Y.5}
\end{equation}
with the initial condition 
\begin{equation}
\mathbf{C}(0) = \mathbf{I}.
\label{Y.6}
\end{equation}

We see that equations (\ref{Y.3}), (\ref{Y.5}) do not depend explicitly on the renormalized time $\tau$, which is the key point of the RG approach. Thus, (\ref{Y.3}) and (\ref{Y.5}) determine a renormalization group parametrized by $\tau$, as it is explained in the end of Section~\ref{secBurg}. 

Let us analyze fixed points of the RG equations. According to (\ref{Y.4}) and (\ref{Y.0}), 
we have the constant solution 
\begin{equation}
\mathbf{C}_* = \mathrm{diag}(0,0,1),
\label{Y.7}
\end{equation}
which is a fixed point attractor of (\ref{Y.5}) with initial condition (\ref{Y.6}). Then, the fixed point solution $\mathbf{U}_*({\bm\xi})$, $P_*({\bm\xi})$ of (\ref{Y.3}) is determined by the equations
\begin{equation}
(\mathbf{A}-\mathbf{I})\mathbf{U}_*
-(\mathbf{A}{\bm\xi}+\mathbf{U}_*)\cdot \nabla_{\xi}
\mathbf{U}_*-\mathbf{C}_*\nabla_{\xi}P_* = 0, \quad 
\nabla_{\xi}\cdot\mathbf{U}_* = 0,
\label{Y.8}
\end{equation}
According to (\ref{Y.2}), the fixed point solution defines 
the self-similar flow  
\begin{equation}
\mathbf{u}(\mathbf{x},t) = t'^{(\mathbf{A}-\mathbf{I})}
\mathbf{U}_*(t'^{-\mathbf{A}}\mathbf{x}'),\quad
p(\mathbf{x},t) 
= t'^{(2a_3-2)}P_*(t'^{-\mathbf{A}}\mathbf{x}'). 
\label{Y.9}
\end{equation}
For each velocity component, the first expression reads
\begin{equation}
u_j(\mathbf{x},t) = t'^{(a_j-1)}U_{*j}(x'_1/t'^{a_1},x'_2/t'^{a_2},x'_3/t'^{a_3}),\quad j = 1,2,3.
\label{Y.10}
\end{equation}

Velocity field and pressure in (\ref{Y.9}) are not exact solutions of the Euler equations (\ref{X.1}), 
since multiple scaling is not a symmetry of the inviscid incompressible flow. 
However, for solutions $\mathbf{U}$ attracted to the fixed point $\mathbf{U}_*$, (\ref{Y.9}) 
yields the asymptotic form of the flow. This asymptotic expression is valid in the vanishingly small 
neighborhood of the blowup, i.e., for
\begin{equation}
x'_1 \sim t'^{a_1},\quad 
x'_2 \sim t'^{a_2},\quad
x'_3 \sim t'^{a_3},\quad 
t' \to 0,
\label{Y.11}
\end{equation}
corresponding to constant values of ${\bm\xi}$.
This fact can also be checked directly by substituting (\ref{Y.9}) into (\ref{X.1}) and noting that the pressure terms of the first two Euler equations  are asymptotically small. 
The asymptotic blowup solution (\ref{Y.9}) is observable, if $(\mathbf{U}_*,P_*,\mathbf{C}_*)$ is an attractor of the RG equations 
(allowing for irrelevant unstable modes related to system symmetries, e.g., space translations and rotations).

The vorticity computed for 
the velocity (\ref{Y.10}) grows as $\omega \sim t'^{(a_1-a_3-1)}$ determined by the dominant derivative $\partial u_1/\partial x_3$. Since $a_3 > a_1$, this agrees with the Beale--Kato--Majda theorem~\cite{beale1984remarks}.
Note that the first two components of the vector field (\ref{Y.10}) satisfy exactly 
the Euler equations with vanishing pressure, i.e., we have
\begin{equation}
\partial u_{1}/\partial t+\mathbf{u}\cdot\nabla u_1 = 0, \quad 
\partial u_{2}/\partial t+\mathbf{u}\cdot\nabla u_2 = 0, \quad 
\nabla \cdot \mathbf{u} = 0,
\label{Y.12}
\end{equation}
as one can check by the substitution of (\ref{Y.10}) into (\ref{Y.12}) and using (\ref{Y.7}), (\ref{Y.8}). 
The Euler equation for the third component remains unchanged and determines the pressure. 
We see that the pressure decouples from system (\ref{Y.12}) for the velocity components. Hence, the inconsistency related to 
determining the pressure function mentioned in the previous section does not appear in the multiple-scale theory. 

\section{Multiple-scale exponential singularity}
\label{secExp2}

In this section we consider extension of the multiple-scale 
RG theory to the case of exponential scaling introduced in Section~\ref{secExp}. 
Let us assume that a flow $\mathbf{u}(\mathbf{x},t)$ with smooth initial condition of finite energy is regular for all $t \ge 0$ and forms a singularity as $t \to \infty$ at the point $\mathbf{x}_b$ with $\mathbf{u}_b = 0$. Exponential renormalization with multiple scales is defined using the diagonal matrices 
\begin{equation}
e^{\mathbf{B}t} 
= \mathrm{diag}\,(e^{b_1t},e^{b_2t},e^{b_3t}),\quad
\mathbf{B} = \mathrm{diag}\,(b_1,b_2,b_3).
\label{V.1}
\end{equation}
We consider the case 
\begin{equation}
0 < b_1 \le b_2 < b_3
\label{V.0}
\end{equation}
with a single dominant exponent $b_3$.
The case of equal $b_2$ and $b_3$ can be analyzed similarly, and 
the single-scale solutions of Section~\ref{secExp} correspond to $b_1 = b_2 = b_3$.

The renormalized solution is defined as
\begin{equation}
\mathbf{u}(\mathbf{x},t) 
= e^{-\mathbf{B}t}\widetilde{\mathbf{U}}({\bm \xi},t),\quad 
p(\mathbf{x},t) 
= e^{-2b_3t}\widetilde{P}({\bm \xi},t),\quad 
{\bm \xi} = e^{\mathbf{B}t}\mathbf{x}'.
\label{V.2}
\end{equation}
The first relation written for each vector component has the form
\begin{equation}
u_j = e^{-b_jt}\widetilde{U}_j({\bm \xi},t),\quad j = 1,2,3,\quad
{\bm\xi} = (x_1/e^{-b_1t},x_2/e^{-b_2t},x_3/e^{-b_3t}).
\label{V.2b}
\end{equation}
Substituting (\ref{V.2}) into (\ref{X.1}) yields the RG equations
\begin{equation}
\widetilde{\mathbf{U}}_t
= \mathbf{B}\widetilde{\mathbf{U}}
-(\mathbf{B}{\bm\xi}+\widetilde{\mathbf{U}})\cdot \nabla_{\xi}
\widetilde{\mathbf{U}}-\widetilde{\mathbf{C}}\nabla_{\xi}\widetilde{P}, \quad 
\nabla_{\xi}\cdot\widetilde{\mathbf{U}} = 0,
\label{V.3}
\end{equation}
where $\widetilde{\mathbf{C}}$ is the diagonal matrix 
\begin{equation}
\widetilde{\mathbf{C}}(t) = e^{-2b_3t}e^{2\mathbf{B}t} = \mathrm{diag}\,(e^{2t(b_1-b_3)},e^{2t(b_2-b_3)},1).
\label{V.4}
\end{equation}
This matrix satisfies 
\begin{equation}
\widetilde{\mathbf{C}}_t 
= 2(\mathbf{B}-b_3\mathbf{I})\widetilde{\mathbf{C}},\quad
\widetilde{\mathbf{C}}(0) = \mathbf{I}.
\label{V.5}
\end{equation}

As in the previous section, we obtained the consistent RG theory, where equations (\ref{V.3}), (\ref{V.5}) do not depend explicitly on time. 
Equations (\ref{V.5}) have the attracting constant solution (\ref{Y.7}). 
Thus, the fixed point solution $\widetilde{\mathbf{U}}_*({\bm\xi})$, 
$\widetilde{P}_*({\bm\xi})$ of (\ref{V.3}) is determined by the equations
\begin{equation}
\mathbf{B}\widetilde{\mathbf{U}}_*
-(\mathbf{B}{\bm\xi}+\widetilde{\mathbf{U}}_*)\cdot \nabla_{\xi}
\widetilde{\mathbf{U}}_*-\mathbf{C}_*\nabla_{\xi}\widetilde{P}_* = 0, \quad 
\nabla_{\xi}\cdot\widetilde{\mathbf{U}}_* = 0.
\label{V.8}
\end{equation}
According to (\ref{V.2}), the fixed point defines the self-similar asymptotic solution  
\begin{equation}
\mathbf{u}(\mathbf{x},t) = e^{-\mathbf{B}t}
\widetilde{\mathbf{U}}_*(e^{\mathbf{B}t}\mathbf{x}'),\quad
p(\mathbf{x},t) 
= e^{-2b_3t}\widetilde{P}_*(e^{\mathbf{B}t}\mathbf{x}'). 
\label{V.9}
\end{equation}
For each velocity component, the first expression reads
\begin{equation}
u_j(\mathbf{x},t) = e^{-b_jt}\widetilde{U}_{*j}(x'_1/e^{-b_1t},x'_2/e^{-b_2t},x'_3/e^{-b_3t}),\quad j = 1,2,3.
\label{V.10}
\end{equation}
The vorticity computed for 
the velocity (\ref{V.10}) grows as $\omega \sim e^{(b_3-b_1)t}$ determined by the dominant derivative $\partial u_1/\partial x_3$. 

Similar to (\ref{Y.9}), velocity field and pressure in (\ref{V.9}) are not exact solutions of the Euler equations (\ref{X.1}), but they provide an asymptotic solution valid in the vanishingly small 
neighborhood of a singularity 
\begin{equation}
x'_1 \sim e^{-b_1t},\quad 
x'_2 \sim e^{-b_2t},\quad
x'_3 \sim e^{-b_3t},\quad 
t \to \infty.
\label{V.11}
\end{equation}
This solution describes an observable singularity, 
if the fixed point $(\widetilde{\mathbf{U}}_*,\widetilde{P}_*,\mathbf{C}_*)$ is 
an attractor of the RG equations (allowing for irrelevant unstable modes related 
to system symmetries).
In this case, (\ref{V.10}) is a universal asymptotic form of a singularity. 
Note that the coefficients $b_1$, $b_2$ and $b_3$ are multiplied by the same 
positive factor under a change of 
time scale and, hence, only the ratios $b_1/b_3$ and $b_2/b_3$ are expected to be universal. 
As in Section~\ref{secExp}, one can obtain the multiple-scale exponential singularity 
expressions as the limit of the formulas for blowup in Section~\ref{secMS} 
with $a_j = b_ja$ and $a \to \infty$. Also,
the asymptotic solution (\ref{V.9}) is the exact solution of (\ref{Y.12}), 
as it can be checked by the substitution using (\ref{V.8}). 

We remark that a scaling of pressure different from (\ref{Y.2}) and (\ref{V.2}) 
can also be considered in the RG scheme. For example, 
the scaling $p(\mathbf{x},t) = \widetilde{P}({\bm\xi},t)$ with $b_1 = b_2 = 1$ was 
suggested in~\cite{brachet1992numerical}. For such scaling, however, 
the problem for fixed points of the RG equations is not well-posed, due to similar 
reasons as described in Section~\ref{secExp}.  

\section{Self-similar exponential singularity in numerical simulations}
\label{secSin}

In several studies, numerical solutions 
were interpreted in favor of finite-time blowup, suggesting the growth of maximum 
vorticity as $\max |\omega| \sim 1/t'$, see, e.g., \cite{kerr1993evidence,pelz2001symmetry}. 
However, this asymptotic relation was not confirmed 
in later computations \cite{hou2006dynamic,grafke2008numerical}. 
Several numerical studies \cite{brachet1992numerical,pumir1990collapsing,grafke2008numerical} 
suggested exponential time dependence of vorticity.
Self-similarity of numerical solutions was discussed 
in \cite{brachet1983small,brachet1992numerical,kerr2005velocity}. 
See also the review of numerical results in \cite{gibbon2008three}.     

\begin{figure}
\centering 
\includegraphics[width=0.4\textwidth]{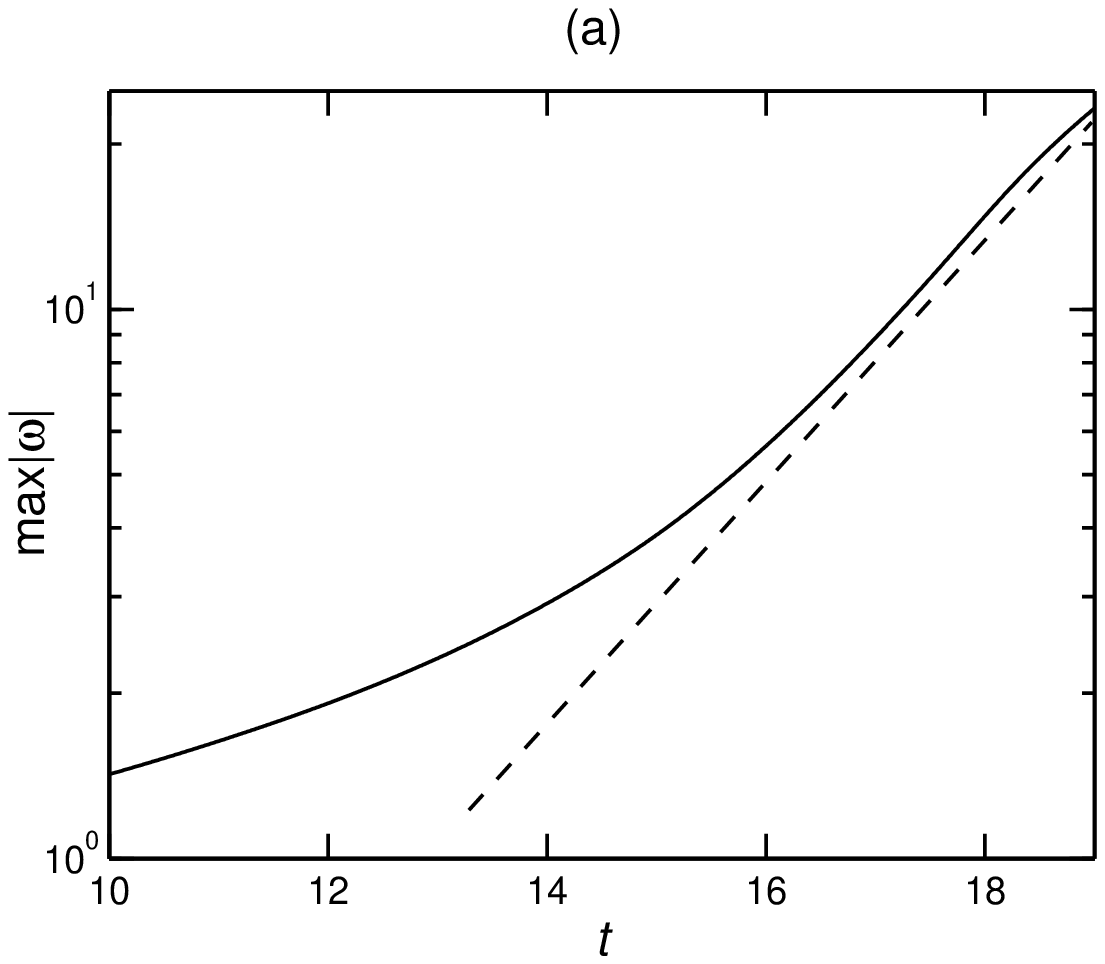}
\hspace{2mm}
\includegraphics[width=0.35\textwidth]{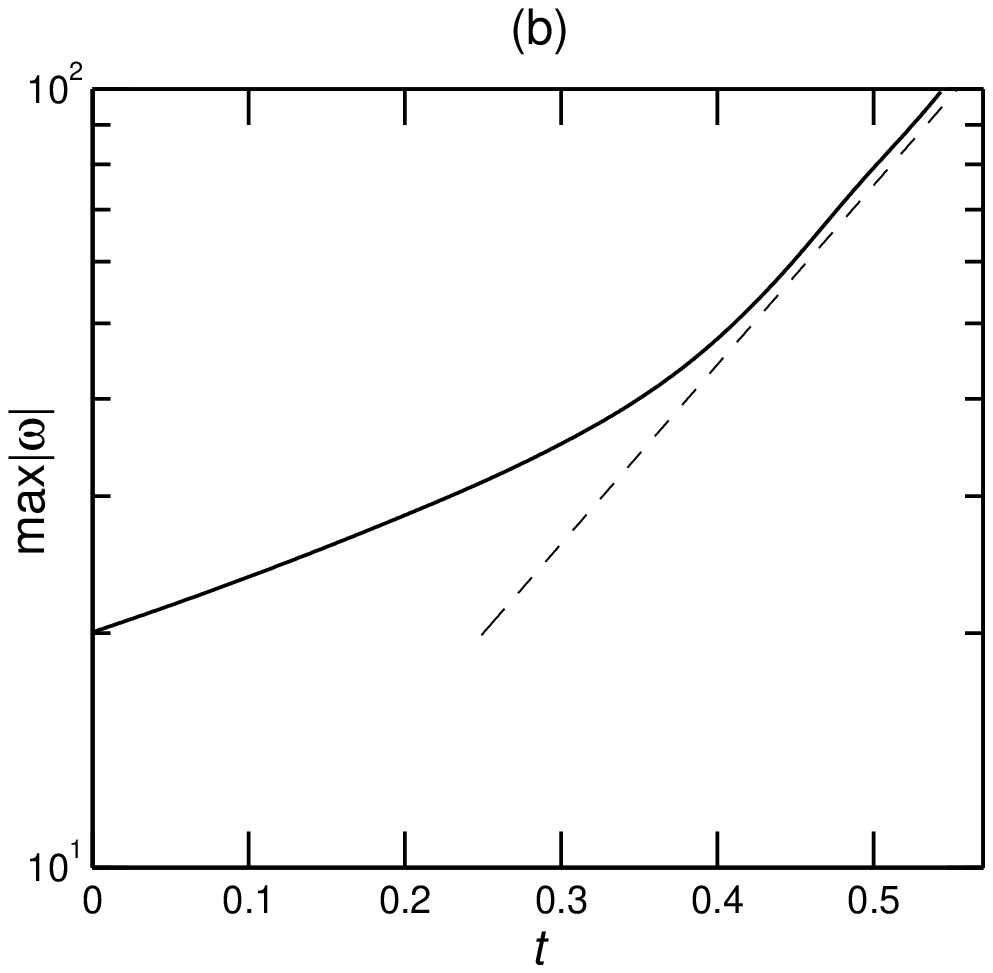}
\caption{Dependence of the maximum vorticity in log-scale on time reconstructed 
from (a) Fig.~9 in \cite{hou2008blowup} and (b) Fig.~1 in \cite{grafke2008numerical}. 
One can notice the asymptotic exponential time dependence (straight dashed lines).}
\label{figHou}
\end{figure}

Fig.~\ref{figHou} shows dependence of maximum vorticity on time 
reconstructed from Fig.~9 in \cite{hou2008blowup} and Fig.~1 in \cite{grafke2008numerical}.
Initial conditions in these simulations have the form of 
antiparallel vortex tubes in \cite{hou2008blowup} and of Kide--Pelz 12 vortices 
in \cite{grafke2008numerical}. Logarithmic scale is used for the maximum vorticity. One can see that $\max|\omega|$ is approximated very well by the exponential time dependence (straight dashed lines) for large times.
Similar exponential behavior was observed along a Lagrangian trajectory passing 
near the singularity, see Fig.~5 in \cite{grafke2008numerical}.

In this section, we use the RG formalism to test the conjecture~\cite{brachet1992numerical} on exponential scaling and self-similarity of flow near a singularity. The appropriate 
self-similar solution is described by (\ref{V.10}). 
For maximum vorticity, it yields 
the asymptotic relation $\max |\omega| \approx \omega_0 e^{(b_3-b_1)t}$ given by the 
dominant term $\partial u_1/\partial x_3$, in agreement with Fig.~\ref{figHou}.

It is convenient to use the Fourier 
transformed form of (\ref{V.10}) as
\begin{equation}
u_j(\mathbf{k},t) = e^{-(b_j+b_1+b_2+b_3) t}
\widetilde{U}_{*j}(k_1/e^{b_1t},k_2/e^{b_2t},k_3/e^{b_3t}),
\quad j = 1,2,3,
\label{eqS2}
\end{equation}
where $\mathbf{u}(\mathbf{k},t)$ is the Fourier transform of $\mathbf{u}(\mathbf{x}_b+\mathbf{x}',t)$
and $\widetilde{\mathbf{U}}_*(\mathbf{k})$ is the Fourier transform 
of $\widetilde{\mathbf{U}}_*({\bm\xi})$. 
The asymptotic relation (\ref{V.10})
is valid in a small neighborhood of a singularity in physical space (\ref{V.11}). Hence, the self-similarity must be observed for
large $k_j$, namely, for
\begin{equation}
k_j \sim e^{b_jt}, \quad j = 1,2,3,\quad t \to \infty.
\label{eqS2asb}
\end{equation}
In particular, it follows from (\ref{eqS2asb}) with the conditions (\ref{V.0}) that, 
for large $t$,
\begin{equation}
k_1 \ll k_3, \quad k_2 \ll k_3,\quad
k = \|\mathbf{k}\| \approx k_3.
\label{eqS2ll}
\end{equation}
This means that the self-similar part of solution 
is concentrated near the axis $k_3$ in the Fourier space. 
Expressions (\ref{eqS2})--(\ref{eqS2ll}) imply  
the self-similar asymptotic behavior of the energy spectrum
\begin{equation}
E(k,t) \approx e^{-b_Et}\widetilde{E}(k/e^{b_3t}), 
\quad k \approx k_3 \sim e^{b_3t},\quad t \to \infty,
\label{eqNN2}
\end{equation}
which scales as the dominant variable $k_3$. The function $\widetilde{E}(k)$ and 
the constant $b_E$ can be expressed in terms of $\widetilde{\mathbf{U}}_*(\mathbf{k})$ and $b_j$.

\begin{figure}
\centering 
\includegraphics[width=0.42\textwidth]{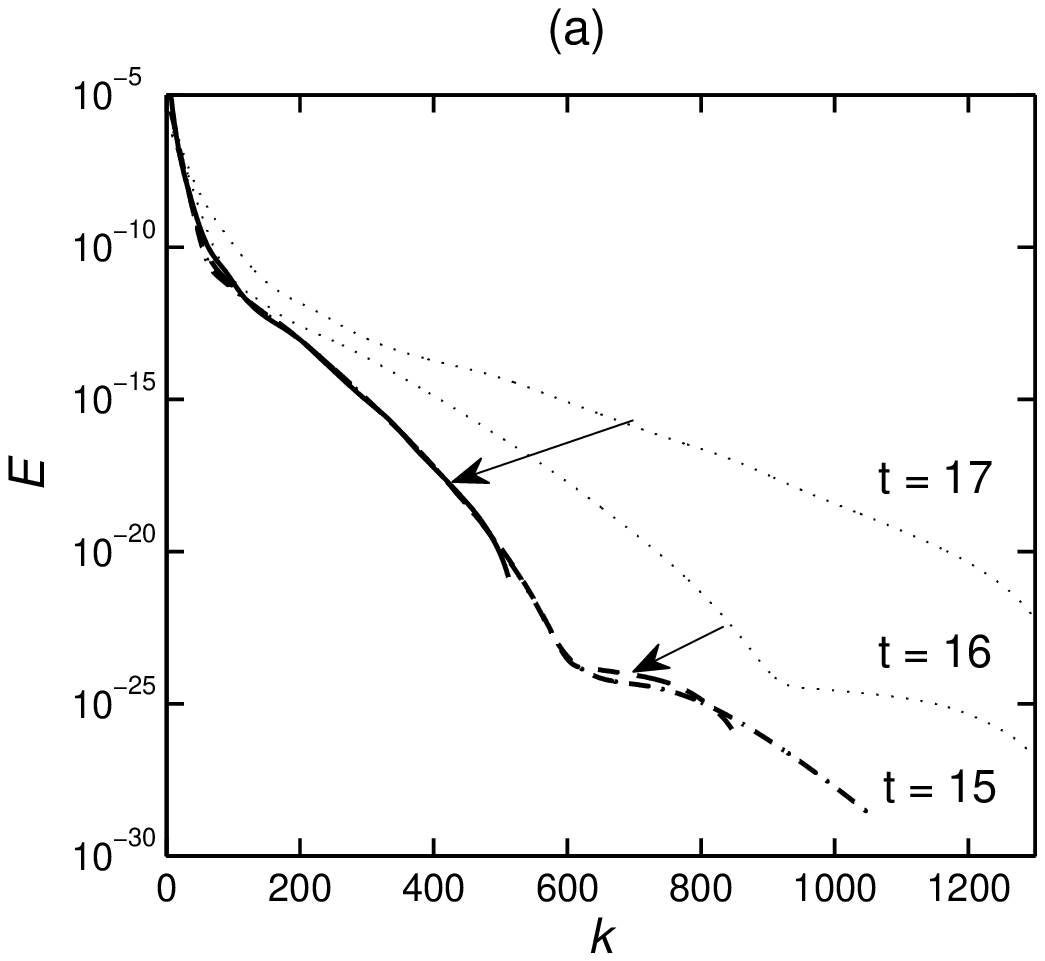}
\includegraphics[width=0.4\textwidth]{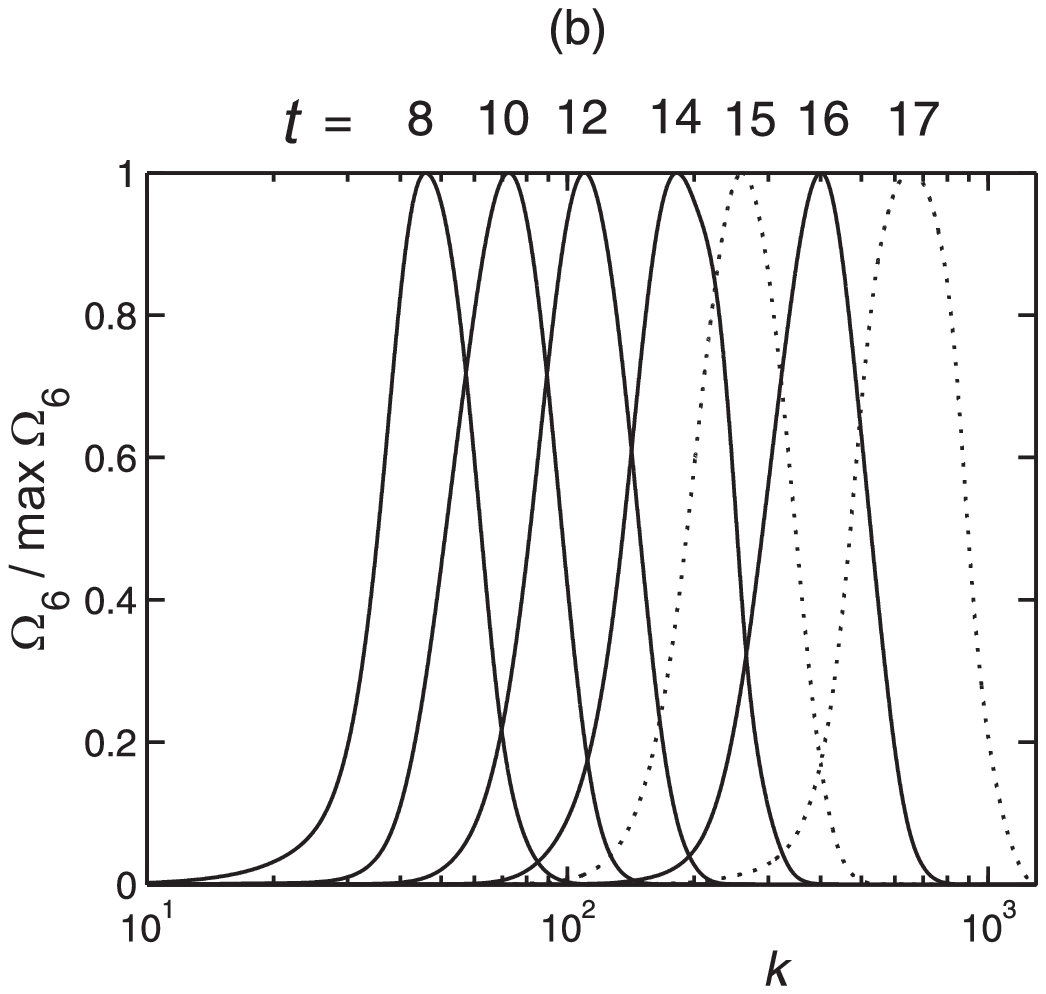}
\includegraphics[width=0.146\textwidth]{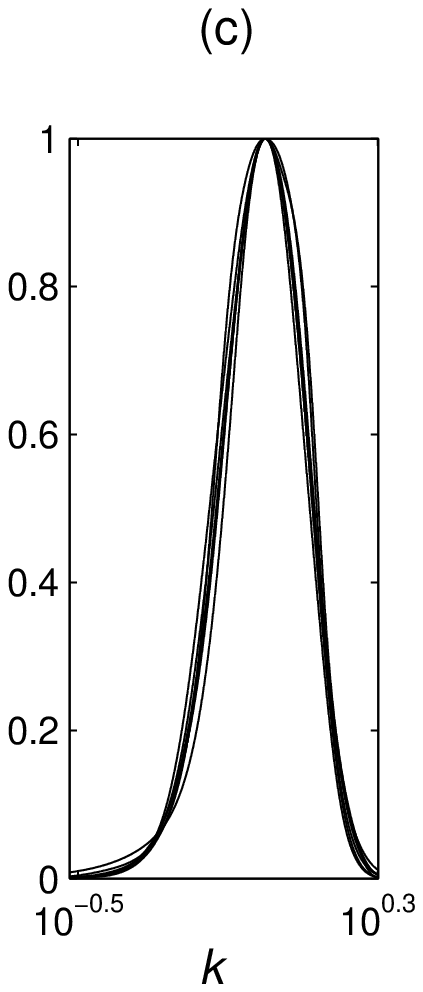}
\caption{Energy spectra reconstructed from 
Figs.~9 and 10 in \cite{hou2007computing}. 
(a) Energy spectra before and after scaling (\ref{eqHL3}). Log-scale is used for $E$. The scaled spectra match in agreement with (\ref{eqNN2}).
(b) Normalized spectra $\Omega_6(k,t)$ at different times with log-scale for $k$ 
demonstrating a ``traveling wave'' 
moving toward large $k$. (c) Profiles of the previous figure shifted to match at the maximum, 
demonstrating self-similar asymptotic form in agreement with (\ref{eqNN3}). }
\label{figHou2}
\end{figure}

Below we verify the self-similarity hypothesis by checking (\ref{eqNN2}) 
for the numerical data reconstructed from Figs.~9 and 10 in \cite{hou2007computing}. 
Fig.~\ref{figHou2}(a) shows the graphs of energy spectra $E(k,t)$ at times $t = 15,16,17$
before and after the scaling chosen as
\begin{equation}
E(k,15),\quad
e^{1.5}E(e^{0.43}k,16),\quad
e^{3}E(e^{0.93}k,17).
\label{eqHL3}
\end{equation}
Good matching of the scaled profiles is observed. 
The scaling exponents change almost linearly in time in agreement with the 
asymptotic expression (\ref{eqNN2}). Similar behavior was observed in~\cite{brachet1983small,brachet1992numerical} based on the approximation $E(k,t) = c(t)k^{-n(t)}e^{-2\delta(t)k}$.

More detailed analysis can be carried out using 
the spectra $\Omega_n(k,t) = k^{2n}E(k,t)$ of generalized enstrophies considered earlier in \cite{brachet1983small}. According to (\ref{eqNN2}), 
these spectra must scale exponentially in time as
\begin{equation}
\Omega_n(k,t) \approx e^{(2nb_3-b_E)t}\widetilde{\Omega}_n(k/e^{b_3t}), 
\quad k \sim e^{b_3t},\quad t \to \infty,
\label{eqNN3}
\end{equation}
where $\widetilde{\Omega}_n(k) = k^{2n}\widetilde{E}(k)$. The reason to consider $\Omega_n(k,t)$ 
instead of $E(k,t)$ is the following. For sufficiently large $n$, the factor $k^{2n}$ 
(corresponding to the spatial mean-square derivative) suppresses 
the energy spectrum in the region of small wave numbers, $k \ll e^{b_3t}$, 
where the asymptotic relation (\ref{eqNN3}) is not valid. 
As a result, the function $\Omega_n(k,t)$ is large for $k \sim e^{b_3t}$, where 
the self-similarity is expected, and gets small far from this region.

\begin{figure}
\centering 
\includegraphics[width=0.47\textwidth]{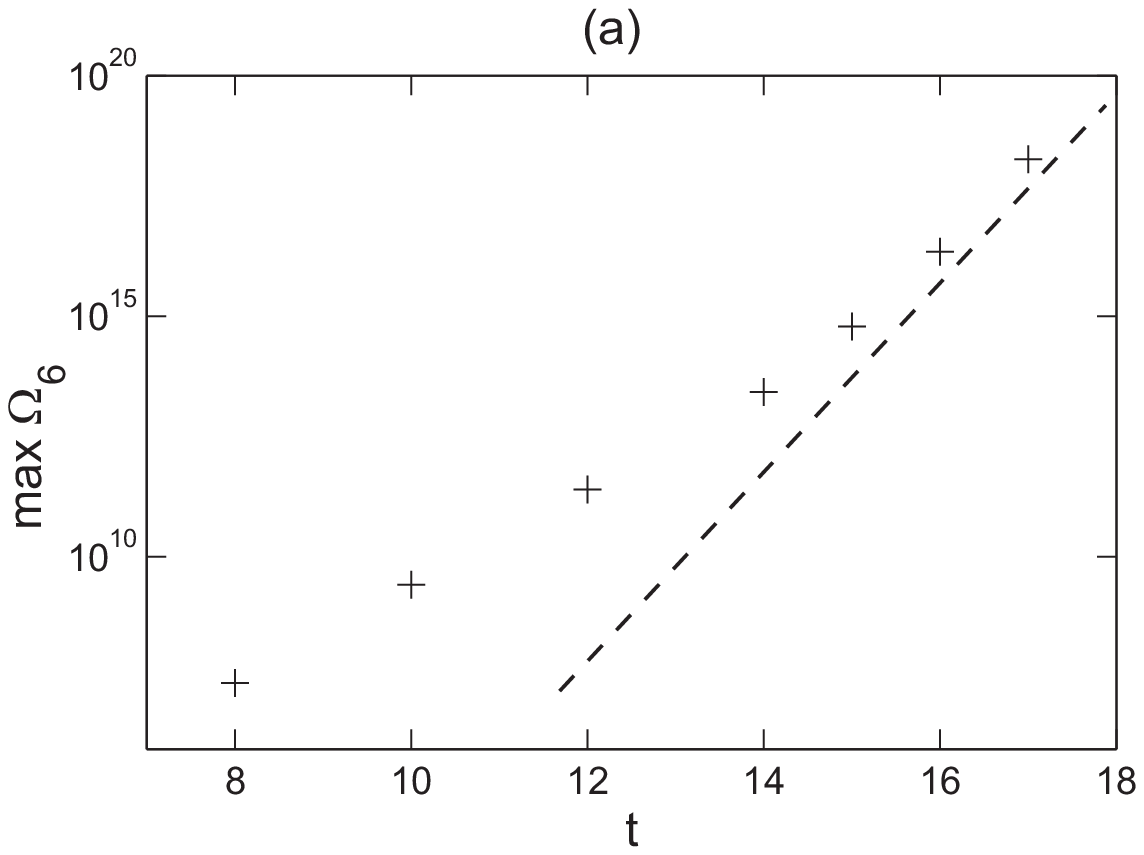}
\includegraphics[width=0.47\textwidth]{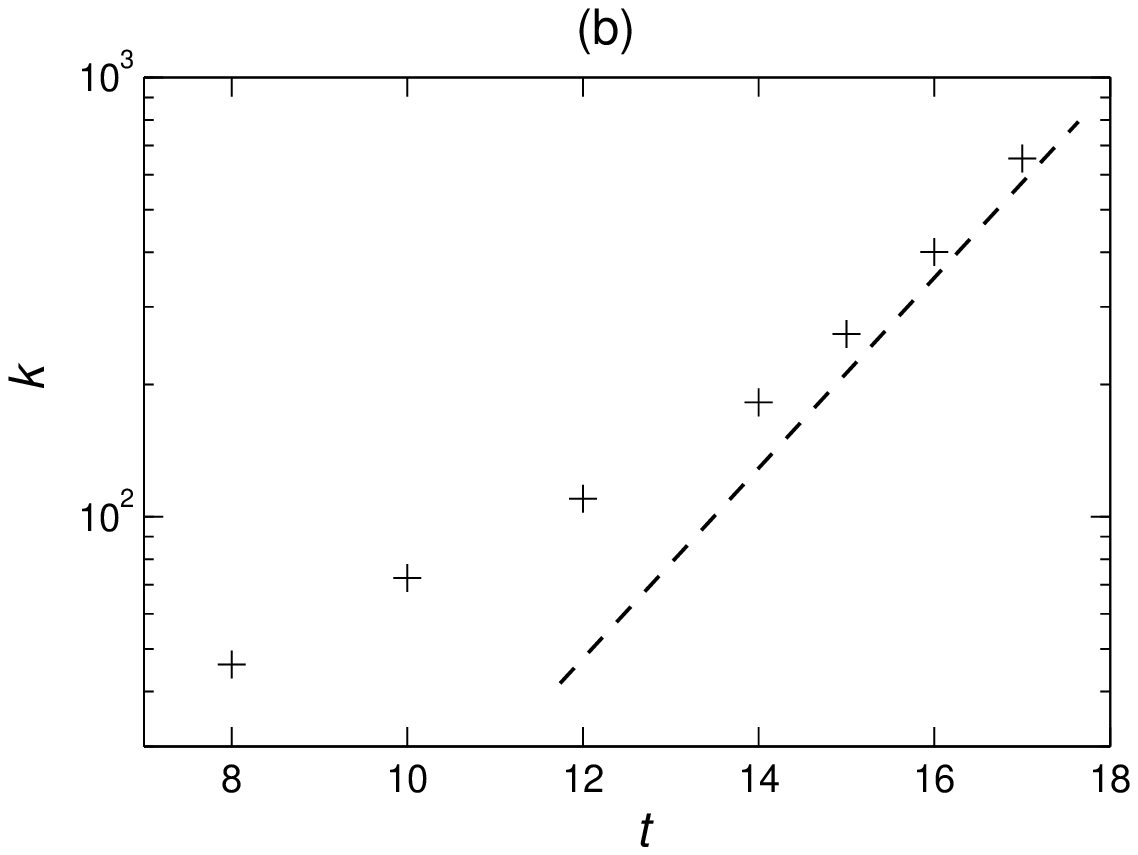}
\caption{Analysis of energy spectra reconstructed from 
Figs.~9 and 10 in \cite{hou2007computing}. Shown are the time dependence of 
(a) $\displaystyle\max_k \Omega_6(k,t)$ and 
(b) the corresponding values of $k$ for the 
profiles in Fig.~\ref{figHou2}(b). Log-scale is used for the vertical coordinate, 
demonstrating asymptotic exponential dependence (dashed lines).}
\label{figHou3}
\end{figure}

Fig.~\ref{figHou2}(b) shows the function $\Omega_6(k,t)$ divided by its maximum value at times 
$t = 8$, $10$, $12$, $14$, $15$, $16$, $17$. The shape of this function 
is almost independent of time, 
as shown in  Fig.~\ref{figHou2}(c), where the profiles are shifted 
to match at the maximum. Fig.~\ref{figHou3} presents the 
values of $\displaystyle \max_k \Omega_6$ and the corresponding values of $k$  
for the profiles in Fig.~\ref{figHou2}(b). The plots demonstrate asymptotic 
exponential dependence for $t \ge 14$ (straight dashed lines in logarithmic scale). 
Fig.~\ref{figHou3} can be compared with the asymptotic exponential growth of maximum vorticity 
in Fig.~\ref{figHou}(a), corresponding to the same numerical simulation.
Note that the blowup in the inviscid Burgers equation (\ref{eq1}) 
as well as in inviscid shell models of turbulence is characterized in the Fourier space 
by the behavior very similar 
to Fig.~\ref{figHou2}(b) under appropriate renormalization, 
see~\cite{dombre1998intermittency,mailybaev2012}.

The presented analysis of numerical results supports the conjecture that 
the inviscid incompressible flow has a singularity developing exponentially in time 
and having asymptotic self-similar structure (\ref{V.10}). 
The RG theory of Section~\ref{secExp2} suggests that the renormalized velocity 
field and ratios of scaling coefficients may be universal, i.e., 
independent of initial conditions. This universality is a powerful criterion, 
which can be checked numerically. Such a test, however, is nontrivial, 
since the universality may be sensitive to symmetry transformations, see Section~\ref{secBurg}.

\section{Conclusion}

The problem of existence and structure of singularities developing 
in finite time (blowup) or infinite time from smooth initial conditions of finite energy
in incompressible Euler equations is considered. 
These singularities may be studied using the renormalization group (RG) approach. 
The central point of this approach is deriving the RG equations, 
which determine the flow evolution combined with renormalization
of space, time and velocities. 
A fixed point attractor of the RG equations, if it exists, 
describes a universal self-similar form of observable singularities.

In this paper, we described extensions of the RG formalism, which include multiple scales and different scaling laws. We explained the possibility of two types of universal self-similar flow structures valid asymptotically in a small neighborhood of a singularity.
The first type describes formation of a finite time singularity (blowup)
with the power law scaling $x'_j \sim t'^{a_j}$, $j = 1,2,3$. In the limit $a_j \to \infty$, 
we obtain the second type corresponding to solutions with exponential 
scaling, $x'_j \sim e^{-b_jt}$, which describe singularities developing exponentially in infinite time. 
Such a limit implies, in particular, that the exponential (infinite time) singularity cannot be distinguished by numerical methods
from the finite-time blowup with very large scaling coefficients $a_j$. 

We showed that numerical results obtained by 
Hou and Li~\cite{hou2007computing,hou2008blowup} and Grafke et al. \cite{grafke2008numerical} support the conjecture of exponential scaling of flow singularity~\cite{brachet1983small,brachet1992numerical}. The analysis shows that the singularity may be described by a universal self-similar solution 
given by the RG theory. The universality provides an effective criterion to be considered in future numerical studies. One should take into account, however, that universality may be sensitive to system symmetries. 

\section*{Acknowledgment} 
The author is grateful to D.S.~Agafontsev and E.A.~Kuznetsov for useful comments. This work was supported by CNPq under grant 477907/2011-3 and CAPES under grant PVE.

\bibliographystyle{unsrt}
\bibliography{refs}

\end{document}